# Halogenation Thermodynamics of Pyrrolidinium-Based Ionic Liquids

Vitaly Chaban

Universidade Federal de São Paulo, São Paulo, Brazil

**Abstract**. Room-temperature ionic liquids (RTILs) exhibit large difference between melting and boiling points. They are highly tunable thanks to numerous accessible combinations of the cation and the anion. On top of that, cations can be functionalized using methods of organic chemistry. This paper reports gas-phase thermodynamics (enthalpy, entropy, Gibbs free energy) of the halogenation reactions (fluorination, chlorination, bromination) involving protonated pyrrolidine $C_4H_{10}N^+$, protic N-ethylpyrrolidinium $C_4H_9N(C_2H_5)^+$, and aprotic N-ethyl-N-methylpyrrolidinium $C_4H_8N(CH_3)(C_2H_5)^+$ cations. Substitution of all symmetrically non-equivalent hydrogen atoms was compared based of the thermodynamic favorability. Fluorination of all sites is much more favorable than chlorination, whereas chlorination is somewhat more favorable than bromination. This is not trivial, since electronegative fluorine and chlorine have to compete for the already insufficient number of electrons with other atoms belonging to the pyrrolidinium-based cations. The difference between different reaction sites within the cations is modest, although it often exceeds kT at simulated temperatures. The correlation between thermodynamics and electronic density distribution has been established, which allows new simple prediction of the reaction pathways. The reported results inspire further chemical modifications of the pyrrolidinium-based RTILs to achieve ever finer tunability of physical chemical properties.

TOC Graphic

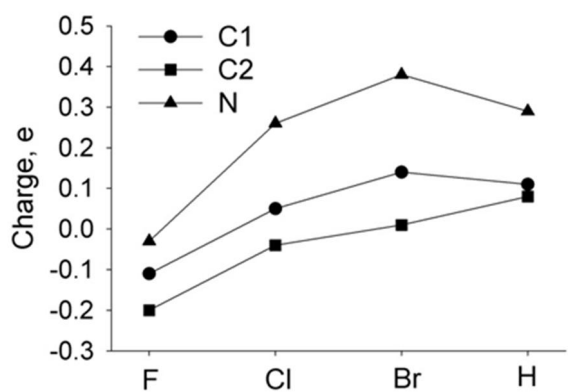 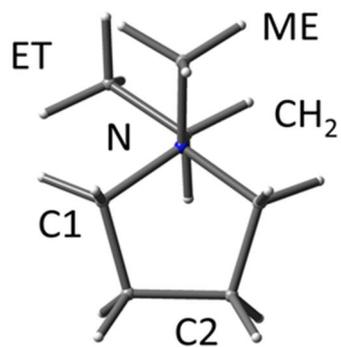

**Introduction**

Room-temperature ionic liquids (RTILs) constitute a large and permanently growing class of chemical compounds, which are based on the organic cation and organic or inorganic anion.[1-15] With just a few exceptions, the cations of RTILs are typically based on a nitrogen containing heterocycle, where the nitrogen atom is primarily responsible for a deficient electron density. It should be understood, however, that many of the discussed cations retain aromatic nature of the corresponding heterocycles, which implies delocalized valence electrons. That is why no atom in the RTIL cation carries a charge of +1e, while most atoms are slightly electron deficient. Together with asymmetry thanks to the grafted hydrocarbon chains, this feature explains why ionic liquids differ from conventional inorganic salts so drastically. Anions of RTILs are less tunable if they are inorganic, such as tetrafluoroborate, hexafluorophosphate, dicyanamide, tricyanomethanide, chloride, bromide, etc. If the anion is organic, it can be modified using the same synthetic approaches as in the case of the cation.

Ionic liquids can be both protic and aprotic. Protic RTILs represent a broad and promising subclass of compounds.[16,17] They have long been used as sources of a highly basic chloride anion for spectroscopic studies and to obtain salt mixtures with low melting temperatures. A large group of protic RTILs with benign and edible cations and biologically interesting anions is available. Ohno and coworkers[18] reported a large number of protic ionic liquids. These researchers used proton transfer from hydrogen bis(trimethylsulfonyl)imide, the conjugate basis of which is an anion, which is known to be responsible for low melting points. In general, melting points and glass transition temperatures appear lower for protic RTILs than for similar aprotic RTILs. Consequently, ionic conductivity of protic RTILs is also higher (consider e.g. ethylammonium nitrate[19,20]), which may be important for their fuel cell applications[21,22] in conjunction with proper cosolvents. Protic RTILs exhibit extensive possibilities to engender hydrogen bonding networks in the condensed phases.

Pyrrolidinine or tetrahydropyrrole, $C_4H_9N$, can be viewed as a cyclic amine with four carbon atoms in the cycle. This is a transparent liquid with a smell of ammonia. The pyrrolidine fragment is widely represented in natural compounds. Pyrrolidine can be protonated in the acidic environments to provide $C_4H_{10}N^+$. $C_4H_{10}N^+$ is a building block for an emerging class of the pyrrolidinium-based RTILs. This charged compound can be stabilized by grafting hydrocarbon chains in place of hydrogen atoms, i.e. $C_4H_8NH_2^+ \rightarrow C_4H_{10}N(CH_3)(C_2H_5)^+$. Coupling of $C_4H_{10}N(CH_3)(C_2H_5)^+$ with a suitable anion provides a new RTIL. Research attention to this class of RTILs is yet insufficient,[23-32] as compared to imidazolium- and pyridinium-based RTILs.[2,10,13,33-35]

Halogenated derivatives are omnipresent in the organic synthesis, both as the end-products and in the numerous established methods for total synthesis. Fluorine, chlorine, and bromine atoms, when substituting hydrogen in the organic molecules, change chemical and physical chemical properties significantly and sometimes drastically.[36-38] Many widely-applied organic solvents were obtained by halogenating aliphatic hydrocarbons (gases). Modification of RTILs using halogenation is interesting for a few reasons: (1) more available solvents with slightly variable sets of properties approach an era of task-specific ionic liquids; (2) halogen atoms are able to block specific binding sites and promote additional binding sites; (3) more electronegative atoms, as compared to hydrogen, induce novel charge density distributions within a cation.

This work reports an accurate gas-phase thermodynamics (enthalpy, entropy, free energy) of fluorination, chlorination and bromination of the three pyrrolidinium-based cations, $C_4H_{10}N^+$, protic N-ethylpyrrolidinium $C_4H_9N(C_2H_5)^+$, and aprotic N-ethyl-N-methylpyrrolidinium $C_4H_8N(CH_3)(C_2H_5)^+$. The below discussion is devoted to atomistic-precision interpretation why some reaction sites are more favorable and other sites are less favorable. The correlation scheme involving partial electrostatic charges has been established, which allows for cheap predicting more preferable halogenation sites without costly free energy calculations.

**Methodology**

Thermodynamics of fluorination, chlorination, and bromination of the N-ethyl-N-methylpyrrolidinium cation was investigated. The direction of the halogenation reaction taking part at constant temperature and constant pressure is determined by an evolution of the Gibbs free energy, $G = f(T,P)$, over reaction stages. The Gibbs free energy is expressed through a linear combination of enthalpy (H) and entropy (S) as $G = H - TS$, where T is an absolute temperature. This paper reports enthalpy, entropy and the Gibbs free energy for the substitution reactions $EM\text{-}PYRR + Hal_2 = EM\text{-}PYR(Hal) + HHal$ between 200 and 500 K.

The results were derived from molecular partition functions using the established equations of statistical mechanics. Wave function optimization followed by geometry optimization, where necessary, was supplemented by computation of force constants and corresponding vibrational frequencies. Vibrational frequencies constitute second derivatives of energy with respect to mass-weighted atomic coordinates. The wave function must be optimized at high level of theory, since electron correlation is responsible for accurate thermodynamic potentials. The coupled-cluster theory[39] was used in this work as an electronic structure method. A calculation employing the coupled-cluster technique starts with a conventional one-electron self-consistent field procedure. Afterwards, correlation between electrons is accounted for. The implementation of the coupled-cluster theory employed in the present work uses single and double substitutions from the Hartree-Fock determinant. Furthermore, it includes triple excitations non-iteratively.[40] The split-valence triple-zeta 6-311G Pople basis set with added polarization and diffuse functions to all atoms was applied. All electrons were considered explicitly. The wave function energy convergence criterion at every self-consistent field step was set to $10^{-8}$ Hartree. The global minimum of each structure was located following the well-established basin hopping algorithm.[41] Partial charges were calculated by fitting electrostatic potential, as described previously.[42] The described electronic structure methods are available in the GAMESS simulation suite.[43]

**Results and Discussion**

We considered three pyrrolidinium-based cations, as depicted in Figure 1, to observe and investigate different halogenation sites. First, the six sites (Figure 2) of the N-ethyl-N-methylpyrrolidinium cation were studied (Tables 1-3). 'C1' stands for the two carbon atoms of the ring bound to the nitrogen, 'N', atom; 'C2' stands for the two carbon atoms of the ring, which are not bound to 'N'; 'ME' and 'ET' are the terminal $sp^3$ carbon atoms of the methyl and ethyl chains, respectively. 'CH$_2$' is the methylene group of the ethyl chain. Second, the methyl chain was removed to obtain the N-ethylpyrrolidinium, a protic cation. The corresponding proton at the 'N' site' is also able to participate in the substitution reactions. Third, the positively charged pyrrolidinium moiety without hydrocarbon chains, $C_4H_{10}N^+$ (protonated pyrrolidine), was considered. The three supposedly different reaction sites, 'C1', 'C2', and 'N', were investigated (Table 4) in $C_4H_{10}N^+$.

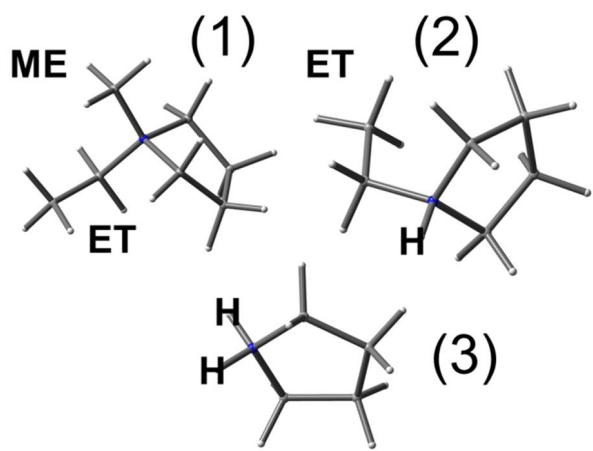

Figure 1. Three pyrrolidinium-based cations investigated in the present work: (1) N-ethyl-N-methylpyrrolidinium $C_4H_8N(CH_3)(C_2H_5)^+$; (2) protic N-ethylpyrrolidinium $C_4H_9N(C_2H_5)^+$; (3) protonated pyrrolidine $C_4H_{10}N^+$.

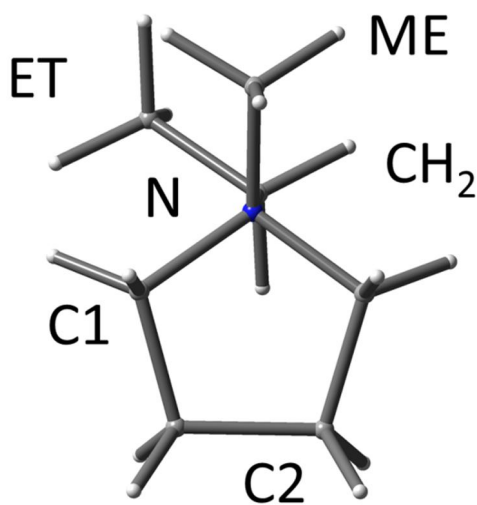

Figure 2. Designation of the reaction sites of the N-ethyl-N-methylpyrrolidinium cation. The reaction sites of the protic N-ethylpyrrolidinium and protonated pyrrolidine use the same names, where applicable.

Gibbs free energy change is responsible for the direction of the process at constant temperature and pressure. A negative Gibbs energy of a reaction does not, however, necessarily mean that a reaction occurs under given conditions, since certain stages can be kinetically forbidden. Nevertheless, a negative Gibbs free energy change does mean that products are more stable than reactants. Consequently, the corresponding reaction can always be performed using this or that transformation pathway. Thermodynamic data are of high importance for synthetic chemistry, because they allow to (1) understand possible products, by-products and intermediates; (2) find out a thermal effect of the reaction; (3) identify the most probable reaction sites in cases where a reactant contains a number of similar reaction sites. This knowledge allows to defend undesirable reaction sites prior to functionalization of an entire molecule.

Simulation of chemical reactions in real-time is impossible at the present stage of development of computational science. Exceptions to this rule are rare and include only reactions with very low activation barriers and very negative free energy alterations. Certain slower reactions can be described using reactive force fields, which are, unfortunately, available for quite a limited set of compounds. Such simulations must be done at high temperatures (1000-3000 K) to make a reaction happen over an observable time interval, usually not exceeding a nanosecond. In this context, thermodynamic analysis brings a lot of important results, which are helpful for chemists.

This work summarizes thermodynamic potentials for a large number of halogenation (fluorination, chlorination and bromination) reactions involving the pyrrolidinium-based cations. Halogenation of ionic liquids is a very interesting transformation, since it allows to achieve significantly different physical chemical properties. For instance, hydrogen bonding, if any, of the cation with the anion or polar molecular cosolvent can be abandoned by elimination of the corresponding hydrogen atom. Furthermore, an electron density can be redistributed fostering formation of new reaction sites. The pyrrolidinium-based cations contain a significant number of possible halogenation sites, 'C1', 'C2', 'N', 'Me', 'ET', 'CH$_2$'. While it can be generally anticipated that halogenation at any of these sites is thermodynamically favorable and it likely becomes less favorable when going from molecular fluorine to molecular bromine as a reactant, a mutual relationship of these sites is unclear. Knowledge and thoughtful prediction of this relationship are important to predict reaction products upon halogenation.

According to the ab initio calculations performed in the present work, fluorination of all cations is significantly more favorable than chlorination, while chlorination is systematically more favorable than bromination (Tables 1-4). Almost all halogenation reactions are thermodynamically allowed. The only two exceptions are bromination of the 'N' site of the N-ethylpyrridinium cation (+39.7 kJ mol$^{-1}$) and bromination of the 'C1' site of the protonated pyrrolidine (+41.1 kJ mol$^{-1}$). Fluorination and chlorination are favorable irrespective of the reaction site. Interestingly, substitution of one hydrogen atom in the –NH$_2$– group by one bromine atom is thermodynamically permitted (-68.6 kJ mol$^{-1}$). One must conclude that the ethyl chain creates a steric hindrance for the corresponding substitution reaction to take place. This information is non-intuitive and important for projecting chemical syntheses.

Table 1. Standard thermodynamic potentials for the fluorination reaction of the N-ethyl-N-methylpyrrolidinium and N-ethylpyrrolidinium (see the 'N' site) cations at standard conditions

| Site | Enthalpy, kJ mol$^{-1}$ | Entropy, J mol$^{-1}$ K$^{-1}$ | Free energy, kJ mol$^{-1}$ |
|---|---|---|---|
| C1 | -462 | -16.6 | -457 |
| C2 | -463 | -18.7 | -458 |
| N | -287 | +8.5 | -289 |
| ME | -446 | -15.7 | -441 |
| CH$_2$ | -467 | -18.0 | -462 |
| ET | -442 | -12.8 | -438 |

Entropic contribution to halogenation is negative in most cases. Decrease of entropy does not favor a direct reaction. It should be kept in mind, however, that –T×S is marginal as compared to the contribution of enthalpy (Figure 3). Compare, the maximum entropic contribution in the case of fluorination amounts to 1.3% (the 'C2' site). This contribution increases somewhat for chlorination (5 to 15%) and additionally increases for bromination (4 to 31%). Nevertheless, in all considered cases the enthalpy change dominates the chemical transformation. The reported results suggest that chlorination and fluorination can be, with a significant accuracy, considered using only the enthalpy change. This observation is important, since computation of enthalpy is systematically cheaper than computation of the Gibbs free energy.

Table 2. Standard thermodynamic potentials for the chlorination reaction of the N-ethyl-N-methylpyrrolidinium and N-ethylpyrrolidinium (see the 'N' site) cations at standard conditions

| Site | Enthalpy, kJ mol$^{-1}$ | Entropy, J mol$^{-1}$ K$^{-1}$ | Free energy, kJ mol$^{-1}$ |
|---|---|---|---|
| C1 | -96.2 | -16.4 | -91.3 |
| C2 | -114 | -17.7 | -109 |
| N | -7.60 | +4.57 | -9.02 |
| ME | -90.2 | -5.24 | -88.6 |
| CH$_2$ | -106 | -19.5 | -99.9 |
| ET | -95.6 | -14.1 | -91.4 |

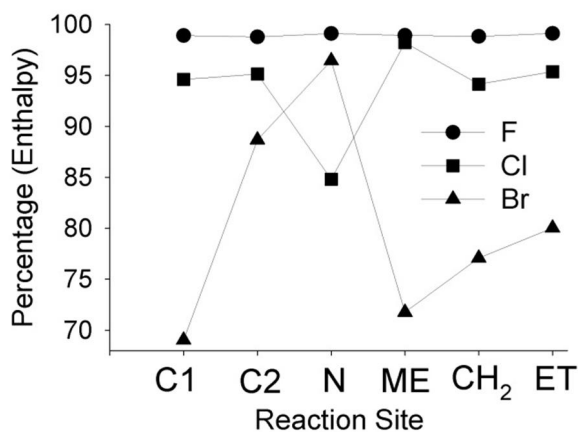

Figure 3. Enthalpy contributions, x(H), to the Gibbs free energy for the halogenation reactions at standard conditions depending on the reaction site. The contributions of entropic factor (T×S) are, consequently, [100 - x(H)]%. Circles correspond to fluorination, squares correspond to chlorination, triangles correspond to bromination.

The observed increase of the entropic impact in the row F < Cl < Br is natural. It correlates well with weakening of the bond strength in the row C-F > C-Cl > C-Br. Entropy must be considered in the case of reactions (physical chemical processes), where an enthalpy alteration is modest. Indeed, the bromination reaction is only slightly thermodynamically favorable. The energy changes (Table 3) exceed the k×T product, however.

Table 3. Standard thermodynamic potentials for the bromination reaction of the N-ethyl-N-methylpyrrolidinium and N-ethylpyrrolidinium (see the 'N' site) cations at standard conditions

| Site | Enthalpy, kJ mol$^{-1}$ | Entropy, J mol$^{-1}$ K$^{-1}$ | Free energy, kJ mol$^{-1}$ |
|---|---|---|---|
| C1 | -22.9 | -18.1 | -17.5 |
| C2 | -43.3 | -14.7 | -38.9 |
| N | +41.1 | +4.71 | +39.7 |
| ME | -21.8 | -15.9 | -16.9 |
| CH$_2$ | -31.7 | -19.7 | -25.8 |
| ET | -24.8 | -13.8 | -20.7 |

It is interesting to note that differences exist between the analogous reaction sites in N-ethyl-N-methylpyrrolidinium, N-ethylpyrrolidinium (protic cation) and protonated pyrrolidine.

For instance, attachment of a halogen atom to the nitrogen atom of the pyrrolidine ring appears clearly more favorable than at the alternative positions if the alkyl chains (methyl and ethyl) are absent. Considering bromination as an example, $[H-N(C_4H_8)-H]^+ + Br_2 = [H-N(C_4H_8)-Br]^+ +$ HBr brings 68.6 kJ mol$^{-1}$. In turn, $[C_2H_5-N(C_4H_8)-H]^+ + Br_2 = [C_2H_5-N(C_4H_8)-Br]^+ +$ HBr costs 39.7 kJ mol$^{-1}$. Reactions at other sites are also thermodynamically different. Fluorination at 'C1' of $C_4H_{10}N^+$ brings only 266 kJ mol$^{-1}$, which is over 1.7 times less than it is brought by fluorination at the same site of N-ethyl-N-methylpyrrolidinium cation (458 kJ mol$^{-1}$). The general conclusion is that halogenation of the carbon atoms of the pyrrolidine ring with the N-grafted hydrocarbon chains is more favorable. Contrary to that, halogenation of the nitrogen atom appears more favorable when the hydrocarbon chains are absent.

Table 4. Standard thermodynamic potentials for fluorination, chlorination, and bromination of $C_4H_{10}N^+$ at standard conditions

| Site | Enthalpy, kJ mol$^{-1}$ | Entropy, J mol$^{-1}$ K$^{-1}$ | Free energy, kJ mol$^{-1}$ |
|---|---|---|---|
| | | Fluorination | |
| C1 | -267 | -3.40 | -266 |
| C2 | -480 | +11.5 | -483 |
| N | -481 | +12.3 | -485 |
| | | Chlorination | |
| C1 | +2.50 | +35.2 | -8.05 |
| C2 | -122.8 | +18.8 | -128 |
| N | -134.6 | +13.1 | -139 |
| | | Bromination | |
| C1 | +46.0 | +16.5 | +41.1 |
| C2 | -29.0 | -10.5 | -25.9 |
| N | -64.2 | +14.7 | -68.6 |

The thermodynamics of the considered reactions is weakly dependent on temperature (Table 5). All isobaric heat capacities, $C_p$, are small by absolute value and their dependence on the reaction site and the reactant is not pronounced. $C_p$ for the 'N' site is larger than $C_p$ for other sites. It decays from fluorine to bromine (154 to 106 J kg$^{-1}$ K$^{-1}$), but this trend is not observed when the values are given per mole. Small $C_p$ indicates an insignificant impact of temperature on halogenation, irrespective of the reaction site. Influence of pressure is not considered, because reactions take place in the condensed phase compressibility of which can be safely neglected.

Table 5. Isobaric heat capacities of the fluorination, chlorination and bromination reactions. These values were derived from the thermodynamic calculations within 200-500 K

| Site | $C_p$ (fluorination) | $C_p$ (chlorination) | $C_p$ (bromination) |
|---|---|---|---|
| | | $J\ mol^{-1}\ K^{-1}$ | |
| 'C1' | 8.32 | 9.60 | 9.09 |
| 'C2' | 8.23 | 8.31 | 8.29 |
| 'N' | 20.4 | 20.7 | 20.5 |
| 'ME' | 6.13 | 7.15 | 6.79 |
| 'CH$_2$' | 8.54 | 9.08 | 8.89 |
| 'ET' | 4.64 | 5.37 | 5.32 |
| | | $J\ kg^{-1}\ K^{-1}$ | |
| 'C1' | 62.9 | 64.6 | 47.1 |
| 'C2' | 62.3 | 55.9 | 42.9 |
| 'N' | 154 | 139 | 106 |
| 'ME' | 46.3 | 48.1 | 35.2 |
| 'CH$_2$' | 64.6 | 61.1 | 46.0 |
| 'ET' | 35.1 | 36.1 | 27.6 |

One of the most important results of this work is that the 'C1', 'C2', 'N', 'ME', 'ET', and 'CH$_2$' reaction sites of the pyrrolidinium-based cations are not thermodynamically equivalent. Although the found difference is inferior to the difference due to different halogen molecules, it still exceeds the k×T product in many cases. This means that certain order of halogenation must be expected. Programmed partial halogenation is possible. The products of the reactions will contain a mixture in which fluorine, chlorine and bromine atoms substitute hydrogen atoms at different positions, both in the pyrrolidine ring and in the side hydrocarbon chains. Such information must be available to synthetic chemists to thoughtfully plan transformations with a particular pyrrolidinium derivative in mind. The discussed modifications of the cations are able to impose significant alterations on the non-bonded interactions within pyrrolidinium-based RTILs and, therefore, to further tune their physical chemical properties.

The reasons why some halogenation sites are more preferable than others may be complicated for understanding, even though the entropic contribution of these reactions is modest. In the following, discussion of the reaction sites is performed in light of partial electrostatic charges. Redistribution of an electron density over the cation upon a chemical reaction is a fundamental factor determining the free energy gain. The partial charges (Figure 4) are, in most cases, enough to understand the recorded trends in thermodynamics (Tables 1-4). Recall that the Pauling-scale electronegativity decays in the row F (4.0) > Cl (3.0) > Br (2.8) > H (2.1). Knowing electronegativity of carbon (2.5) is also important, since all these atoms are bound to carbon in the pyrrolidinium-based cations. Fluorine is always negatively charged, chlorine is nearly neutral, $|q| > 0.1e$, bromine is always electron deficient, even though its electronegativity exceeds that of carbon. This represents a good illustration that electronegativity as a physical property is of predominantly qualitative nature, which suits best for binary compounds.

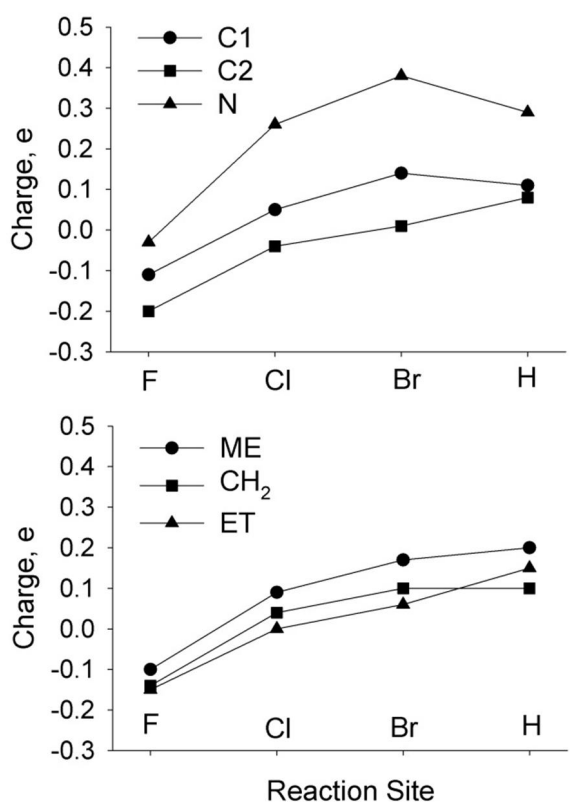

Figure 4. Electrostatic charges of the atoms (F, Cl, Br, H) participating in the halogenation reactions in their respective positions in the pyrrolidinium-based cations.

The partial electrostatic charges help to understand relations between free energies of reactions. For instance, substitution of the hydrogen atom by a halogen atom at 'C1' is somewhat less favorable, as compared to 'C2'. According to Figure 4, the halogen atoms are allowed to maintain a more negative charge at this site. This is in line with their higher electronegativity and, therefore, the corresponding reaction sites are more energetically favorable. The 'C2' reaction site is most distant from the center of the positive charge of the pyrrolidinium-based cations. Therefore, it owns more electron density, which is, upon a reaction, transferred to fluorine, chlorine and bromine in each respective case. Within the N-ethylpyrrolidinium cation (protic), the 'N' site is least favorable, although the halogenation reaction at this site is energetically permitted in the temperature range between 200 and 500 K and atmospheric pressure, except the case of bromination. Indeed, the analysis of partial charges shows that halogen atom must be electron-deficient at this position. Even the fluorine atom is nearly neutral in $[(C_4H_9)(C_2H_5)N-F]^+$, whereas chlorine and bromine atoms are clearly positively charged. Interestingly, Table 3 shows a different trend for the case of protonated pyrrolidine. This is again in concordance with the proposed correlation scheme, since the nitrogen atom is negatively charged, q=-0.18e, in $C_4H_{10}N^+$. The overall positive charge of this ion is achieved due to the two hydrogen atoms, q(H)=+0.28e each, which are bound to the nitrogen atom. When one of these hydrogen atoms of substituted by the methyl chain, a significant amount of the positive charge appear at ME. That is why ME is the least preferable site in the N-ethyl-N-methylpyrrolidinium cation. Many similar correlations can be found using Tables 1-4 and Figure 4, which are left as an exercise for an interested reader.

**Concluding Remarks**

Thermodynamics of fluorination, chlorination and bromination is reported and discussed. It was found that halogenation of the N-ethyl-N-methylpyrrolidinium, N-ethylpyrrolidinium and protonated pyrrolidine cations is thermodynamically allowed in almost all cases. Fluorination is systematically more favorable then chlorination. In turn, chlorination is somewhat more favorable than bromination. This trend originates from the strength of C-F, C-Cl and C-Br polar covalent bonds.

It was found that more favorable reaction sites in the case of the substitution halogenation reactions permit halogen atoms to be electron-richer, which is in concordance with the widely-used electronegativity scale. In principle, this result allows to predict an order of prospective halogenation sites using only an electrostatic charge distribution within a pristine organic cation. Efforts on further chemical modification of ionic liquids must benefit from this observation, since it fosters understanding of the resulting mixture of products and stipulate synthetic solutions to adjust its composition.

**Acknowledgments.** This project was partially funded by CAPES (Brazil).

**Contact Information.** E-mail for correspondence: vvchaban@gmail.com. Tel: +55 12 3309-9573; Fax: +55 12 3921-8857.